\documentclass[letterpaper]{article} 
\usepackage{aaai24}  
\usepackage{times}  
\usepackage{helvet}  
\usepackage{courier}  
\usepackage[hyphens]{url}  
\usepackage{graphicx} 
\usepackage{natbib}  
\usepackage{caption} 
\usepackage{algorithm}
\usepackage{algorithmic}

\usepackage{newfloat}
\usepackage{listings}

\usepackage{multirow}
\usepackage{multicol}
\usepackage{booktabs}
\usepackage{amsmath}
\usepackage{amsfonts}
\usepackage{color}

\DeclareCaptionStyle{ruled}{labelfont=normalfont,labelsep=colon,strut=off} 
\floatstyle{ruled}
\newfloat{listing}{tb}{lst}{}
\floatname{listing}{Listing}
\title{Generalizable Sleep Staging via Multi-Level Domain Alignment}
\author{
	Jiquan Wang\textsuperscript{\rm 1,\rm 2},
	Sha Zhao\textsuperscript{\rm 1,\rm 2}\thanks{Corresponding authors.},
	Haiteng Jiang\textsuperscript{\rm 3,\rm 4,\rm 1},
	Shijian Li\textsuperscript{\rm 1,\rm 2},
	Tao Li\textsuperscript{\rm 3,\rm 4,\rm 1},
	Gang Pan\textsuperscript{\rm 1,\rm 2,\rm 4}\footnotemark[1]
}
\affiliations{
	\textsuperscript{\rm 1}State Key Laboratory of Brain-machine Intelligence, Zhejiang University, Hangzhou, China\\
	\textsuperscript{\rm 2}College of Computer Science and Technology, Zhejiang University, Hangzhou, China\\
	\textsuperscript{\rm 3}Department of Neurobiology, Affiliated Mental Health Center \& Hangzhou Seventh People's Hospital, Zhejiang University School of Medicine, Hangzhou, China\\
	\textsuperscript{\rm 4}MOE Frontier Science Center for Brain Science and Brain-machine Integration, Zhejiang University, Hangzhou, China\\
	\{wangjiquan, szhao, h.jiang, shijianli, litaozjusc, gpan\}@zju.edu.cn
}

\usepackage{bibentry}

\begin{document}
	
	\maketitle
	
	\begin{abstract}
		Automatic sleep staging is essential for sleep assessment and disorder diagnosis. Most existing methods depend on one specific dataset and are limited to be generalized to other unseen datasets, for which the training data and testing data are from the same dataset. In this paper, we introduce domain generalization into automatic sleep staging and propose the task of generalizable sleep staging which aims to improve the model generalization ability to unseen datasets. Inspired by existing domain generalization methods, we adopt the feature alignment idea and propose a framework called SleepDG to solve it. Considering both of local salient features and sequential features are important for sleep staging, we propose a Multi-level Feature Alignment combining epoch-level and sequence-level feature alignment to learn domain-invariant feature representations. Specifically, we design an Epoch-level Feature Alignment to align the feature distribution of each single sleep epoch among different domains, and a Sequence-level Feature Alignment to minimize the discrepancy of sequential features among different domains. SleepDG is validated on five public datasets, achieving the state-of-the-art performance.
		
	\end{abstract}
	
	\section{Introduction}
	
	Sleep plays an important role in human health. 
	Sleep staging refers to the classification of sleep into different sleep stages, which is crucial to identifying sleep problems and other disorders in humans~\citep{he2018effects}.
	Clinically, sleep stages are scored by doctors or experts using electrical activity recorded from sensors attached to different parts of the body.
	A set of signals from these sensors is called a polysomnogram (PSG), consisting of multiple physiological signals recorded, such as electroencephalogram (EEG) and electrooculogram (EOG).
	According to the American Academy of Sleep Medicine (AASM) sleep standard~\citep{Iber2007TheAA}, PSG is usually segmented into 30-second epochs, which are manually classified into five different sleep stages (Wake, N1, N2, N3, and REM) by experts. 

	 In recent years, with the development of deep learning techniques~\citep{sun2016remembered, pan2018rapid, yi2021hippocampal, Wang2023ARO}, many deep learning models~\citep{phan2022automatic} have been proposed to solve the task of automatic sleep staging.
	They are implemented with different network structures based on sequence-to-sequence framework, obtaining good performance in sleep staging from PSG recordings.
	However, most of existing methods adopt the intra-dataset scheme, where the training data and testing data are from the same dataset, ignoring the discrepancies between various datasets and making it difficult to be generalized to unseen sleep staging datasets well~\citep{anido2022analysis}.
	The discrepancies among multiple datasets could be caused by many factors, such as heterogeneous patient population,  different signal channel, different data collection equipment types or manners, or different medical environments. 
	In clinic, a sleep staging model should better be generalized to unseen datasets of any new populations in any environment.
	Some methods~\citep{yoo2021transferring, fan2022unsupervised, wang2022automatic} illustrate that the training and testing samples are not drawn from the same probability distribution which cause \textbf{performance deterioration} when directly applying models trained on source datasets to target datasets. They adopt the ideas of Domain Adaptation (DA) to solve it, but the generalization problem still exist because DA methods assume that the target sleep staging database is accessible.
	
	In order to solve the generalizing-to-unseen-database problem of automatic sleep staging, we introduce the idea of \textbf{domain generalization (DG)} into the sleep staging task:
	A \textit{domain} is composed of data that is sampled from a join distribution of input space and label space.
	The goal of domain generalization is to learn a model from one or several different but related domains (i.e., diverse training datasets) that will perform well on unseen testing domains \citep{wang2022generalizing}.
	The unseen domain is \textit{inaccessible} in training procedure.
	Inspired by some studies \citep{koh2021wilds, robey2021model} exploring hospital-level DG in medical imaging,
	we set a single sleep staging dataset (usually from one hospital) as a \textit{domain} and set the discrepancies between different datasets as \textit{domain shift}.
	Then multiple datasets used for training can be regarded as \textit{source domains} and the dataset used for testing can be seen as \textit{target domain} or \textit{unseen domain}.
	Our goal is to learn a model from several source domains (several sleep datasets) that can perform well at the unseen domain (unseen sleep dataset). We name this task as \textbf{generalizable sleep staging}.
	
	There have been some existing DG studies \citep{muandet2013domain, li2018domain, li2018dg, matsuura2020domain, zhou2020domain} in other applications, which mainly focus on learning domain-invariant feature representations by sample-level feature distribution alignment to solve the sample-level classification problem, referring to mapping single sample into the corresponding label.
	However, different from sample-level classification problem, sleep staging is a sequence-to-sequence classification problem, which maps a sequence of samples into the corresponding sequence of labels.
	According to AASM standard~\citep{Iber2007TheAA}, not only local salient waveforms within each epoch but also transition patterns of sleep stages between neighbor epochs play critical roles in sleep staging. Meantime, there have been some studies~\citep{Eldele2021AnAD, Phan2021XSleepNetMS} proving the importance of both local intra-epoch features and global inter-epoch features for sleep staging.
	Therefore, it is critical to combine intra-epoch feature (epoch-level) alignment and inter-epoch feature (sequence-level) alignment in alignment process, to solve the generalization problem of sleep staging.
	
	In this paper, we introduce DG to sleep staging and propose a framework, SleepDG, to solve the generalization problem of sleep staging.
	Our contributions are as follows:
	\begin{itemize}
		\item 
		We introduce a novel task of generalizable sleep staging to solve the generalization problem of sleep staging in clinic. For the task, we propose \textbf{SleepDG, a DG-based deep learning framework,} which learns domain-invariant feature representations from different  PSG datasets and can perform well on unseen datasets.
		
		\item
		We propose a \textbf{Multi-level Feature Alignment} method consisting of \textbf{Epoch-level Feature Alignment} and \textbf{Sequence-level Feature Alignment} designed to align feature distribution within each single epoch and between epochs in a sleep sequence among different PSG datasets. 
		
		\item
		SleepDG is validated on \textbf{five public datasets} in the DG scenarios, achieving the state-of-the-art performance.
	\end{itemize}

	\section{Related Work}
	\subsection{Automatic Sleep Staging}
	Automatic sleep staging~\citep{melek2021automatic} refers to the classification of sleep epochs into different sleep stages in a sequence-to-sequence fashion.
	Existing deep learning methods almost use a local extractor for epoch features and a global extractor for sequential context features.
	For example, \citet{Supratak2017DeepSleepNetAM, Supratak2020TinySleepNetAE} utilized CNN to extract local features and Bi-LSTM to encode temporal information.
	\citet{Qu2020ARB, Eldele2021AnAD, wang2023narcolepsy} utilized CNN to capture intra-epoch features and the multi-head self-attention to model global temporal context.
	\citet{Perslev2019UTimeAF} utilized a fully-CNN Encoder-Decoder architecture to model local salient wave characteristic and sleep transitional rules. 
	\citet{Phan2019SeqSleepNetEH} proposed a hierarchical RNN named SeqSleepNet to extract local and global features from time-frequency images of multimodal PSG data.
	\citet{Phan2022SleepTransformerAS} proposed SleepTransformer which encodes both local features and temporal features via fully Transformers.
	\citet{Jia2021SalientSleepNetMS} proposed SalientSleepNet, which uses a fully CNN based on the U$^2$-Net to detect the salient waves from multimodal PSG signals and a Multi-scale Extraction Module to capture sleep transition rules.
	Above methods achieve a high performance on sleep staging on specific datasets.
	However, it is difficult for them to be generalized to other unseen datasets.
	
	\subsection{Domain Generalization}
	DG~\citep{gulrajani2021in, zhou2021domain, wang2022generalizing} considers the scenarios where there are domain shift among different domains, which is different from domain adaptation (DA) \citep{wang2018deep, zhao2021plug} because target domain in DG is inaccessible.
	The DG problem was formally introduced by \citet{blanchard2011generalizing, muandet2013domain} as a machine learning problem.
	Learning domain-invariant feature representations by minimizing the feature distribution divergence is a common and effective method for domain generalization.
	Some studies minimized the feature distribution divergence by minimizing the maximum mean discrepancy (MMD) \citep{tzeng2014deep, wang2018visual, wang2020transfer}, second order correlation \citep{sun2016return, sun2016deep}, Wasserstein distance \citep{zhou2020domain} and Similarity Metric \citep{dou2019domain} in deep neural network.
	Some studies minimized the feature distribution divergence by adopting Domain-adversarial neural network (DANN)~\citep{li2018domain} for learning domain-invariant features, which was originally proposed by \citet{ganin2015unsupervised} for DA.
	Most of above methods learn domain-invariant features by minimizing single-sample features divergence, neglecting the crucial role of inter-epoch features in sleep staging, which needs to be improved.	
%

	\begin{figure*}[tb]
		\centering
		\small
		\includegraphics[width=0.8\textwidth]{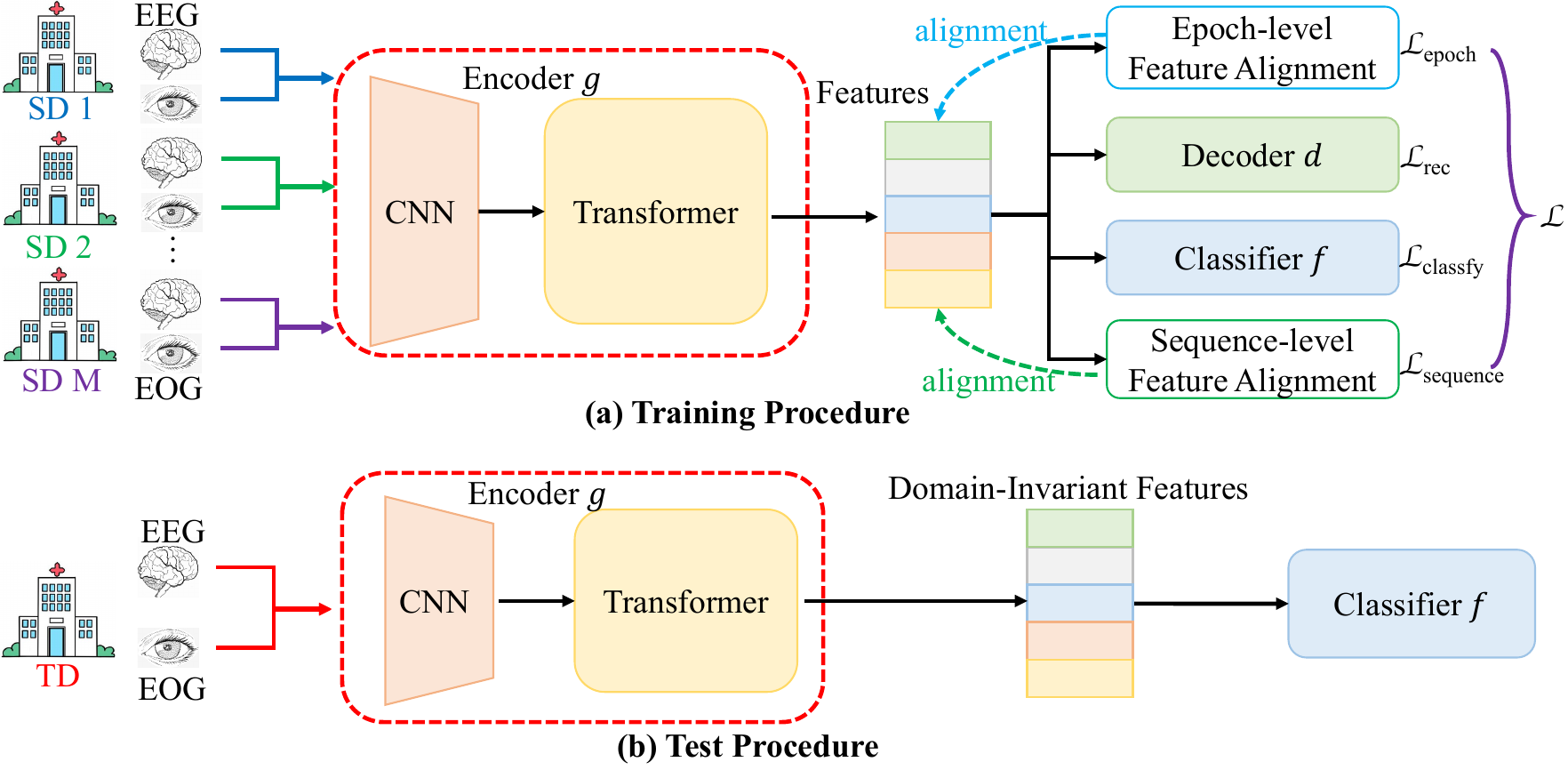}
		\caption{SleepDG overview. Here, SD is source domain and TD is target domain.}
		\label{fig:model}
	\end{figure*}
	
	\section{Problem Formulation}
	We introduce a novel task, called generalizable sleep staging, which is a multi-source DG problem.
	Let $\mathcal{X}$ be the input space and $\mathcal{Y}$ the target space, a \textit{domain} is defined as data sampled from a join distribution $P_{XY}$ on $\mathcal{X} \times \mathcal{Y}$.
	Here, \textbf{we treat one single sleep dataset as one domain}.
	We have several labeled source datasets $D^{S_i}=(X^{S_i}, Y^{S_i})$, where $i \in [1, 2, 3,..., M]$ denotes the $i$-th source domain and $M=4$ in this work.
	$\mathrm{\mathbf{x}}^{S_i}_j = \{x^{S_i}_{j, 1}, x^{S_i}_{j, 2}, ... , x^{S_i}_{j, L}\}$ is the $j$-th  sequence composed of $L$ sleep epochs and $\mathrm{\mathbf{y}}^{S_i}_j = \{y^{S_i}_{j, 1}, y^{S_i}_{j, 2}, ... , y^{S_i}_{j, L}\}$ is the $j$-th sequence composed of corresponding $L$ sleep stages.
	$x^{S_i}_{j, k} \in \mathbb{R}^{n \times C}$, $n$ denotes the number of sampling points in a epoch and $C$ is the number of channels.
	$y^{S_i}_{j, k} \in \{0, 1\}^N$ and $N=5$ denote the number of \textit{sleep stages (Wake, N1, N2, N3, REM)}.
	We also have an unseen target dataset $D^T=(X^T, Y^T)$. 
	In the domain shift scenarios, we supposed the different source datasets $(X^{S_i}, Y^{S_i})$ and target dataset $(X^T, Y^T)$ are sampled from different distributions.
	
	Our generalizable sleep staging task is defined as learning a mapping function $h:\mathcal{X} \rightarrow \mathcal{Y}$ using only source datasets so that the predictive error on an unseen target dataset is minimized. 
	We decompose the prediction function $h$ as $h = f \circ g$, where $g$ is a feature encoder and $f$ is a classifier.
	Therefore, the goal of minimizing the predictive error on an unseen datasets domain can be formulated as:
	\begin{equation}
		\label{eq:goal}	
		\small
		\min_{f,g} \mathbb{E}_{(\mathrm{\mathbf{x}}, \mathrm{\mathbf{y}}) \sim P(X^T, Y^T)} \mathcal{L}(f(g(\mathrm{\mathbf{x}})), \mathrm{\mathbf{y}}) + \lambda \mathcal{L}_{\mathrm{reg}}
	\end{equation}
	where $\mathcal{L}_{\mathrm{reg}}$ denotes some regularization term and $\lambda$ is the tradeoff parameter.
	
	\section{Method}
	\subsection{Overview}

	Our generalizable sleep staging framework SleepDG is summarized in Fig.~\ref{fig:model}.
	In the training phase shown in Fig.~\ref{fig:model}(a), we take the source data $X^{S_i}$ which is from $M$ different domain as inputs and use encoder $g$, decoder $d$, classifier $f$ and Multi-level Feature Alignment to learn domain-invariant feature representations $H^{S_i}$.
	The whole training process is trained in an end-to-end fashion with the supervision of the groundtruth sleep stages.
	As shown in Fig.~\ref{fig:model}(b), in the test phase we take the target data $X^T$ which is from the target domain as input and extract the feature representations $H^T$ by the feature encoder $g$.
	Then we feed $H^T$ into the classifier $f$ to predict the sleep stages $Y^T$.
	
	\subsection{AE-based Feature Encoding} 
	
	In order to learn feature representations $H^{S_i}$, we design a sequence-to-sequence autoencoder (AE), consisting of encoder $g$ and decoder $d$.
	We take the source data $X^{S_i}$ as inputs and extract the feature representations $H^{S_i}$ by encoder $g$.
	The process of features extracting can be expressed as $H^{S_i}=g(X^{S_i})$,
	where $H^{S_i} = \{\mathrm{\mathbf{h}}^{S_i}_j\}^{N^{S_i}}_{j=1}$ is composed of many \textit{sequences}  $\mathrm{\mathbf{h}}^{S_i}_j$, $\mathrm{\mathbf{h}}^{S_i}_j = \{h^{S_i}_{j, 1}, h^{S_i}_{j, 2}, ... , h^{S_i}_{j, L}\}$ is composed of $L$ feature representations which are learned from $L$ sleep epochs, $h^{S_i}_{j, k} \in \mathbb{R}^{d}$ and $d$ denotes the feature dimension.
	Specifically, the feature encoder $g$ consists of CNN and Transformer, where CNN is used to extract intra-epoch features and Transformer is used to extract inter-epoch features.
	We use a decoder $d$ to reconstruct $X^{S_i}$ from $H^{S_i}$.
	The process of reconstruction can be expressed as $\hat{X}^{S_i}=d(H^{S_i})$,
	where $\hat{X}^{S_i}$ is the reconstructed sleep data corresponding to $X^{S_i}$.
	Then, we use the Mean Squared Error (MSE) loss function as the reconstruction loss:
	\begin{equation}
		\label{predicting}
		\small
		\mathcal{L}_{\mathrm{rec}} = \frac{1}{L}\sum^{L}_{k=1} \Vert x^{S_i}_{j, k} - \hat{x}^{S_i}_{j, k} \Vert_2,
	\end{equation}
	where $x^{S_i}_{j, k}$ is a sleep epoch, $\hat{x}^{S_i}_{j, k}$ is the reconstructed sleep epoch and  $\Vert \cdot \Vert_2$ denotes the squared norm.
	
	\subsection{Multi-level Feature Alignment} 
	The feature alignment refers to mapping source data from different datasets into one shared feature space in which $\mathrm{P}(H^{S_i}, Y^{S_i}) = \mathrm{P}(H^{S_j}, Y^{S_j})$ if $i \neq j$.
	We can decompose joint distribution $\mathrm{P}(H^{S_i}, Y^{S_i})$ into a marginal distribution and a conditional distribution by $\mathrm{P}(H^{S_i}, Y^{S_i}) = \mathrm{P}(H^{S_i})\mathrm{P}(Y^{S_i}|H^{S_i})$.
	According to AASM standard \citep{Iber2007TheAA}, sleep experts adopt a uniform rules for manual sleep staging, so we assume that the conditional distributions are the same in different sleep staging datasets.
	In other words, we make a prior assumption that $\mathrm{P}(Y^{S_i}|H^{S_i})=\mathrm{P}(Y^{S_j}|H^{S_j})$, if $i \neq j$.
	Under such a prior assumption, we only need to align the marginal distribution $\mathrm{P}(H^{S_i})$.
	
	Sleep staging is a sequence-to-sequence classification and both local intra-epoch features and global context features among sequential epochs are important for sleep staging~\citep{Eldele2021AnAD, Phan2021XSleepNetMS}. Inspired by this, we design a Multi-level Feature Alignment method consisting of Epoch-level Feature Alignment and Sequence-level Feature Alignment.
	The alignment of sequence features $\mathrm{\mathbf{h}}^{S_i}_j = \{h^{S_i}_{j, 1}, h^{S_i}_{j, 2}, ... , h^{S_i}_{j, L}\}$ could be divided to reducing marginal distribution on single epoch features and conditional distribution between epochs in a sleep sequence:
	\begin{equation}
		\label{diveded}
		\small
		\begin{split}
			\mathrm{P}(\mathrm{\mathbf{h}}^{S_i}_j)&=\mathrm{P}(h^{S_i}_{j, k})\mathrm{P}(\mathrm{\mathbf{h}}^{S_i}_j \setminus \{h^{S_i}_{j, k}\}|h^{S_i}_{j, k})\\
			&= \frac{1}{L}\sum^{L}_{k=1}\mathrm{P}(h^{S_i}_{j, k})\mathrm{P}(\mathrm{\mathbf{h}}^{S_i}_j \setminus \{h^{S_i}_{j, k}\}|h^{S_i}_{j, k}),
		\end{split}
	\end{equation}
	where $\mathrm{P}(h^{S_i}_{j, k})$ is the marginal distribution of each epoch feature and $\mathrm{P}(\mathrm{\mathbf{h}}^{S_i}_j\setminus \{h^{S_i}_{j, k}\}|h^{S_i}_{j, k})$ is the conditional distribution between each epoch feature and other epoch features in a sleep sequence.
	To summarize, \textbf{the feature alignment could be converted to the alignment of $\mathrm{P}(h^{S_i}_{j, k})$ and $\mathrm{P}(\mathrm{\mathbf{h}}^{S_i}_j\setminus \{h^{S_i}_{j, k}\}|h^{S_i}_{j, k})$, which can be called Multi-level Feature Alignment.}
	We call the alignment of $\mathrm{P}(h^{S_i}_{j, k})$ as \textit{Epoch-level Feature Alignment} and the alignment of  $\mathrm{P}(\mathrm{\mathbf{h}}^{S_i}_j\setminus \{h^{S_i}_{j, k}\}|h^{S_i}_{j, k})$ as \textit{Sequence-level Feature Alignment}.
	
	\subsubsection{Epoch-level Feature Alignment}
	Some existing methods minimize the feature distribution divergence by minimizing the maximum mean discrepancy (MMD) \citep{tzeng2014deep, wang2018visual, wang2020transfer} and other methods minimize the domain discrepancy by minimizing the second order correlation \citep{sun2016return, sun2016deep}.
	Inspired by above, we utilize both the first-order statistics (expectation) and the second-order statistics (covariances) as the measure of distribution to minimize the domain discrepancy.
	We set $F^i$ as the set of all features in the source domain $S_i$, so any $h^{S_i}_{j, k} \in F^i$.
	The first-order statistics discrepancy is computed through the following equation:
	\begin{equation}
		\label{eq:first}
		\small
		\mathcal{L}_{\mathrm{first}} = \sum_{i \neq j}\Vert \mathrm{E}(F^i) - \mathrm{E}(F^j) \Vert_2,
	\end{equation}
	where $\mathrm{E}(\cdot)$ is the expectation and $\Vert \cdot \Vert_2$ denotes the squared norm.
	
	The second-order statistics discrepancy is computed through the following equation:
	\begin{equation}
		\label{eq:second}
		\small
		\mathcal{L}_{\mathrm{second}} = \sum_{i \neq j}\Vert \mathrm{COV}(F^i) - \mathrm{COV}(F^j) \Vert^2_F,
	\end{equation}
	where $\mathrm{COV}(\cdot)$ is the covariances matrix and $\Vert \cdot \Vert^2_F$ denotes the squared matrix Frobenius norm.
	
	Combining the first-order statistics discrepancy and the second-order statistics discrepancy, we get the epoch-level discrepancy computed equation:
	\begin{equation}
		\label{eq:epoch}
		\small
		\mathcal{L}_{\mathrm{epoch}} = \mathcal{L}_{\mathrm{first}} + \mathcal{L}_{\mathrm{second}},
	\end{equation}
	By minimizing $\mathcal{L}_{\mathrm{epoch}}$, we can minimize the epoch-level domain discrepancy to align the marginal distribution  $\mathrm{P}(h^{S_i}_{j, k})$.
	
	\subsubsection{Sequence-level Feature Alignment}
	Our Sequence-level Feature Alignment is to align the conditional distribution $\mathrm{P}(\mathrm{\mathbf{h}}^{S_i}_j\setminus \{h^{S_i}_{j, k}\}|h^{S_i}_{j, k})$.
	However, it is very difficult to align this distribution directly.
	In order to simplify this problem, we regard the conditional distribution $\mathrm{P}(\mathrm{\mathbf{h}}^{S_i}_j\setminus \{h^{S_i}_{j, k}\}|h^{S_i}_{j, k})$ as the relationship between the feature of one epoch and the features of other epochs in the sleep sequence.
	Thus, we decide to use the Pearson correlation coefficients to represent the relationship between different epochs in a sequence and to align the expectation of the inter-epoch Pearson correlation coefficients in the sleep sequence from different source sleep datasets.

	We set $R^i_j$ as the correlation matrix of sleep sequence $\mathrm{\mathbf{h}}^{S_i}_j$, and $\rho_{k, t}$ is the element of the $k$-th column and the $t$-th row in $R^i_j$.
	$\rho_{k, t}$ is computed through the following equation:
	\begin{equation}
		\small
		\rho_{k, t}=\frac{Cov(h^{S_i}_{j, k}, h^{S_i}_{j, t})}{\sqrt{Var(h^{S_i}_{j, k})Var(h^{S_i}_{j, t})}},
	\end{equation}
	where $Cov(\cdot)$ is the covariances and $Var(\cdot)$ is the variance.
	After getting $R^i_j$, we can compute the expectation $R^i$ of correlation matrix of source domain $S_i$ through the following equation:
	\begin{equation}
		\small
		R^i = \frac{1}{N^{S_i}} \sum_{j=1}^{N^{S_i}} R^i_j,
	\end{equation}
	where $N^{S_i}$ denotes the number of sleep sequences in a source domain $D^{S_i}$.
	
	Finally, we can compute the sequence-level discrepancy through the following equation:
	\begin{equation}
		\label{eq:sequence}
		\small
		\mathcal{L}_{\mathrm{sequence}} =  \sum_{i \neq j}\Vert R^i - R^j \Vert^2_F,
	\end{equation}
	where $\Vert \cdot \Vert^2_F$ denotes the squared matrix Frobenius norm.
	By minimizing $\mathcal{L}_{\mathrm{sequence}}$, we can indirectly align the conditional distribution $\mathrm{P}(\mathrm{\mathbf{h}}^{S_i}_j\setminus \{h^{S_i}_{j, k}\}|h^{S_i}_{j, k})$ to minimize the sequence-level domain discrepancy. 
	
	\subsection{Training}
	
	
	We design a classifier $f$ to classify each epoch in a sequence into different sleep stages. The classifier $f$ is a fully connected layer with a softmax function.
	We use the cross-entropy (CE) function as the loss function for the sleep staging task:
	\begin{equation}
		\label{eq:classify}
		\small
		\mathcal L_{\mathrm{classify}} =-\sum^{L}_{k=1} \sum^{N}_{l=1} y^{S_i}_{j, k, l}log(\hat{y}^{S_i}_{j, k, l}),
	\end{equation}
	where $y^{S_i}_{j, k, l}$, the $l$-th element of $y^{S_i}_{j, k}$, denotes the probability that $x^{S_i}_{j, k}$ actually belongs to the $l$-th stage, and $\hat{y}^{S_i}_{j, k, l}$, the $l$-th element of $\hat{y}^{S_i}_{j, k}$, denotes the probability that $x^{S_i}_{j, k}$ is predicted to the $l$-th stage.
	Finally, we combine the loss functions for feature alignment and classifier:
	\begin{equation}
		\label{eq:loss}
		\small
		\mathcal L = \mathcal L_{\mathrm{classify}} + \lambda_1 \mathcal{L}_{\mathrm{rec}} + \lambda_2 \mathcal{L}_{\mathrm{epoch}} + \lambda_3 \mathcal{L}_{\mathrm{sequence}},
	\end{equation}
	where $\mathcal L$ is the final loss function used in Eq.~(\ref{eq:goal}), $\lambda_1$,$\lambda_2$ and $\lambda_3$ is the coefficients.
	The training procedure in one epoch can be summarized as Algorithm~\ref{alg:tasks}.
	
	\begin{table*}[tb]
		\centering
		\small
		\begin{tabular}{lccccc} 
			\toprule
			Dataset&Sample Frequency &Recordings&Recordings&EEG channel&EOG channel\\
			&(Hz)&(all)&(we choose)&(we choose)&(we choose)\\
			\midrule
			I. SleepEDFx&100&197&197&Fpz-Cz&horizontal\\
			\textcolor{red}{II. HMC}&256&151&151&F4-M1&E1-M2\\
			\textcolor{red}{III. ISRUC}&200&126&126&F4-M1&E1-M2\\
			\textcolor{red}{IV. SHHS}&125&5,793&150&C4-M1&ROC-LOC\\
			\textcolor{red}{V. P2018}&200&995&150&C3-M2&E1-M2\\
			\bottomrule
		\end{tabular}
		\caption{Summary of the datasets used in our experiments. \textcolor{red}{There is an error in the dataset numbering in the formal version of the AAAI paper; we correct it here.}}\label{tab:datasets}
	\end{table*}
	
	\begin{table*}[tb]
		\centering
		\small
		\begin{tabular}{lcccccccccccc} 
			\toprule
			\textbf{SD}& \multicolumn{2}{c}{II, III, IV, V} & \multicolumn{2}{c}{I, III, IV, V} & \multicolumn{2}{c}{I, II, IV, V} & \multicolumn{2}{c}{I, II, III, V} & \multicolumn{2}{c}{I, II, III, IV}& \multicolumn{2}{c}{\multirow{2}{*}{Avg}}\\
			\cmidrule(lr){1-1}\cmidrule(lr){2-3}\cmidrule(lr){4-5}\cmidrule(lr){6-7}\cmidrule(lr){8-9}\cmidrule(lr){10-11}
			\textbf{TD}&\multicolumn{2}{c}{I} & \multicolumn{2}{c}{II} & \multicolumn{2}{c}{III} & \multicolumn{2}{c}{IV} & \multicolumn{2}{c}{V}&&\\
			\cmidrule(lr){1-1}\cmidrule(lr){2-3}\cmidrule(lr){4-5}\cmidrule(lr){6-7}\cmidrule(lr){8-9}\cmidrule(lr){10-11}\cmidrule(lr){12-13}
			Metrics& ACC & MF1 & ACC & MF1 & ACC & MF1 & ACC & MF1 & ACC & MF1 & ACC & MF1 \\
			\midrule
			DeepSleepNet&72.28&65.72&64.04&62.84&69.71&67.80&64.73&52.79&66.62&60.51&67.48&61.93\\
			U-Time&72.51&65.84&65.13&63.71&70.58&68.10&64.53&51.68&67.35&61.07&67.82&62.28\\
			AttnSleep&73.76&66.93&64.49&62.19&70.39&68.19&64.07&51.82&66.19&60.78&67.78&61.98\\
			ResnetMHA&73.01&65.89&65.18&63.02&70.11&67.54&65.16&53.99&67.89&61.87&68.27&62.46\\
			TinySleepNet&73.34&66.10&65.87&64.01&69.18&66.99&65.76&54.56&68.29&61.36&68.49&62.60\\
			EnhancingCE&73.51&66.69&65.98&64.37&70.56&69.12&65.88&54.39&67.84&61.19&68.75&63.14\\
			SalientSleepNet&73.92&67.59&65.44&63.22&69.93&68.16&66.36&55.49&68.89&63.13&68.91&63.52\\
			\midrule
			SleepDG&\textbf{77.44}&\textbf{71.29}&\textbf{73.85}&\textbf{71.16}&\textbf{78.69}&\textbf{74.44}&\textbf{70.45}&\textbf{60.89}&\textbf{74.74}&\textbf{70.43}&\textbf{75.03}&\textbf{69.64}\\
			\bottomrule
		\end{tabular}
		\caption{Performance comparison with existing non-DG methods for sleep staging. \textcolor{red}{I is SleepEDFx, II is HMC, III is ISRUC, IV is SHHS, V is P2018.}}\label{tab:CompSS}
	\end{table*}
	
	\section{Experiments}
	\subsection{Datasets and Preprocessing}
	
	We evaluated the performance of SleepDG and other methods on \textbf{five different public sleep staging datasets:}
	\textbf{I. SleepEDFx} \citep{Kemp2000AnalysisOA, Goldberger2000PhysioBankPA} consists of 197 PSG recordings, all of which were used for evaluation.
	\textbf{II. HMS} \citep{alvarez2021inter} includes 151 whole-night PSG recordings, and we used all the recordings in the experiments.
	\textbf{III. ISRUC} \citep{khalighi2016isruc} consists of 126 whole-night PSG recordings, all of which were employed in this work.
	\textbf{IV. SHHS} \citep{zhang2018national, quan1997sleep} consists of 5,791 PSG recordings.
	To keep the balance with other datasets, we used the first 150 recordings in our experiments.
	\textbf{V. P2018} \citep{ghassemi2018you} consists of 944 whole-night PSG recordings.
	We employed the first 150 recordings in this work for the balance with other datasets.
	The summary of all the datasets is shown in Tab.~\ref{tab:datasets}.

	In order to ensure that SleepDG can adapt to datasets with different channels, we only selected single-channel EEG and EOG from each dataset.
	
	SleepEDFx and SHHS were scored according to R\&K standard \citep{Wolpert1969AMO}, including W, N1, N2, N3, N4 and REM.
	We merged the N3 and N4 stages into a single stage N3 according to the latest AASM standard \citep{Iber2007TheAA}. 
	Besides, for SleepEDFx, we only kept the PSGs starting from 30 minutes before to 30 minutes after the first and last non-wake epochs as recommended by \citet{Supratak2017DeepSleepNetAM}.
	All the sleep recordings used in our experiments were band-pass filtered (0.3Hz--35Hz), resampled to 100Hz and normalized according to the Z-score standardization.

	\subsection{Settings}
	\textbf{We take turns to select four ones from the five datasets as the source domains for training and set the left one as the unseen domain for testing, where the training data and testing data are from different datasets.}
	We adopt \textit{training-domain validation} \citep{wang2022generalizing} as the strategy of model selection, where each source domain is split into the training part and the validation part.
	Notably, \textbf{the unseen domain is inaccessible in the training procedure}.
	
	We implemented SleepDG based on the PyTorch.
	The source code is publicly available\footnote{\url{https://github.com/wjq-learning/SleepDG}}.
	The model is trained using the Adam optimizer with default settings, the learning rate is set to 1e-3 and the weight decay is set to 1e-4.
	The coefficients $\lambda_1$, $\lambda_2$ and $\lambda_3$ are all set to 0.5.
	The training epoch is 50, the mini-batch size is set to 32 and the dropout rate is 0.1.
	We set the length of sleep epoch sequence as $L=20$ and the feature dimension as $d=512$.
	Accuracy (ACC) and Macro-F1 score (MF1) are used as evaluation metrics.
	We trained the model on one machine with Intel Core i9 10900K CPU and eight NVIDIA RTX 3080 GPUs.
	
	\subsection{Results and Analysis}
	
	\subsubsection{Compared with non-DG Methods for Sleep Staging}
	
	\begin{table*}[tb]
		\centering
		\small
		\begin{tabular}{lcccccccccccc} 
			\toprule
			\textbf{SD}& \multicolumn{2}{c}{II, III, IV, V} & \multicolumn{2}{c}{I, III, IV, V} & \multicolumn{2}{c}{I, II, IV, V} & \multicolumn{2}{c}{I, II, III, V} & \multicolumn{2}{c}{I, II, III, IV}& \multicolumn{2}{c}{\multirow{2}{*}{Avg}}\\
			\cmidrule(lr){1-1}\cmidrule(lr){2-3}\cmidrule(lr){4-5}\cmidrule(lr){6-7}\cmidrule(lr){8-9}\cmidrule(lr){10-11}
			\textbf{TD}&\multicolumn{2}{c}{I} & \multicolumn{2}{c}{II} & \multicolumn{2}{c}{III} & \multicolumn{2}{c}{IV} & \multicolumn{2}{c}{V}&&\\
			\cmidrule(lr){1-1}\cmidrule(lr){2-3}\cmidrule(lr){4-5}\cmidrule(lr){6-7}\cmidrule(lr){8-9}\cmidrule(lr){10-11}\cmidrule(lr){12-13}
			Metrics& ACC & MF1 & ACC & MF1 & ACC & MF1 & ACC & MF1 & ACC & MF1 & ACC & MF1 \\
			\midrule
			BASE&73.89&67.02&65.63&63.85&70.34&68.49&65.10&54.47&68.83&62.85&68.76&63.34\\
			MMD&75.41&70.06&69.39&66.84&75.09&70.60&68.09&57.68&70.71&65.00&71.74&66.04\\
			CORAL&75.38&70.70&70.06&68.57&74.83&72.18&69.45&58.33&71.62&67.81&72.27&67.52\\
			IRM&75.27&70.21&70.24&68.42&74.84&71.81&68.48&56.43&71.60&67.88&72.09&66.95\\
			DANN&74.71&69.56&69.30&67.69&72.69&70.35&67.49&56.17&71.01&67.40&71.12&66.23\\
			\midrule
			SleepDG&\textbf{77.44}&\textbf{71.29}&\textbf{73.85}&\textbf{71.16}&\textbf{78.69}&\textbf{74.44}&\textbf{70.45}&\textbf{60.89}&\textbf{74.74}&\textbf{70.43}&\textbf{75.03}&\textbf{69.64}\\
			\bottomrule
		\end{tabular}
		\caption{Performance comparison with existing DG methods for sleep staging. \textcolor{red}{I is SleepEDFx, II is HMC, III is ISRUC, IV is SHHS, V is P2018.}}	\label{tab:CompDG}
	\end{table*}
	
	\begin{table*}[tb]
		\centering
		\small
		\begin{tabular}{lcccccccccccc} 
			\toprule
			\textbf{SD}& \multicolumn{2}{c}{II, III, IV, V} & \multicolumn{2}{c}{I, III, IV, V} & \multicolumn{2}{c}{I, II, IV, V} & \multicolumn{2}{c}{I, II, III, V} & \multicolumn{2}{c}{I, II, III, IV}& \multicolumn{2}{c}{\multirow{2}{*}{Avg}}\\
			\cmidrule(lr){1-1}\cmidrule(lr){2-3}\cmidrule(lr){4-5}\cmidrule(lr){6-7}\cmidrule(lr){8-9}\cmidrule(lr){10-11}
			\textbf{TD}&\multicolumn{2}{c}{I} & \multicolumn{2}{c}{II} & \multicolumn{2}{c}{III} & \multicolumn{2}{c}{IV} & \multicolumn{2}{c}{V}&&\\
			\cmidrule(lr){1-1}\cmidrule(lr){2-3}\cmidrule(lr){4-5}\cmidrule(lr){6-7}\cmidrule(lr){8-9}\cmidrule(lr){10-11}\cmidrule(lr){12-13}
			Metrics& ACC & MF1 & ACC & MF1 & ACC & MF1 & ACC & MF1 & ACC & MF1 & ACC & MF1 \\
			\midrule
			BASE&73.89&67.02&65.63&63.85&70.34&68.49&65.10&54.47&68.83&62.85&68.76&63.34\\
			AE&74.34&68.08&66.84&64.82&72.21&70.67&65.76&55.38&69.48&63.22&69.73(\textbf{+0.97})&64.43(\textbf{+1.09})\\
			EA&75.88&71.00&71.56&69.69&75.63&72.10&66.98&56.67&71.62&67.61&72.33(\textbf{+3.57})&67.41(\textbf{+4.07})\\
			SA&74.41&68.56&69.47&68.84&72.09&69.60&68.98&58.51&70.71&67.00&71.13(\textbf{+2.37})&66.50(\textbf{+3.16})\\
			AE+EA&75.89&70.90&72.34&70.18&76.58&73.23&68.28&57.60&72.81&68.34&73.18(\textbf{+4.42})&68.05(\textbf{+4.71})\\
			AE+SA&74.38&70.16&70.30&69.13&75.65&72.99&67.23&55.25&71.74&68.44&71.86(\textbf{+3.10})&67.19(\textbf{+3.85})\\
			\midrule
			SleepDG&\textbf{77.44}&\textbf{71.29}&\textbf{73.85}&\textbf{71.16}&\textbf{78.69}&\textbf{74.44}&\textbf{70.45}&\textbf{60.89}&\textbf{74.74}&\textbf{70.43}&\textbf{75.03(+6.27)}&\textbf{69.64(+6.30)}\\
			\bottomrule
		\end{tabular}
		\caption{Performance comparison with ablated methods. \textcolor{red}{I is SleepEDFx, II is HMC, III is ISRUC, IV is SHHS, V is P2018.}}\label{tab:AblationStudy}
	\end{table*}
	
	Firstly, we compared with several non-DG methods for automatic sleep staging:
	\textbf{DeepSleepNet} \citep{Supratak2017DeepSleepNetAM} is a classical CNN-BiLSTM network for extracting local features and learning transition rules.
	\textbf{U-Time} \citep{Perslev2019UTimeAF} is
	a fully-CNN encoder-decoder architecture for time series segmentation applied to sleep staging.
	\textbf{AttnSleep} \citep{Eldele2021AnAD}
	is composed of a multi-resolution CNN  and a multi-head self-attention with causal convolutions.
	\textbf{ResnetMHA} \citep{Qu2020ARB}
	uses a residual CNN to capture local features and a self-attention to model global temporal context.
	\textbf{TinySleepNet} \citep{Supratak2020TinySleepNetAE}
	is a classical model based on CNN and RNN, with a smaller number of model parameters.
	\textbf{EnhancingCE} \citep{Phyo2022EnhancingCE}
	captures both of local salient and global contextual features via two auxiliary tasks.
	\textbf{SalientSleepNet} \citep{Jia2021SalientSleepNetMS}
	proposes a fully CNN based on the U$^2$-Net to detect multimodal salient waves.
	
	We implemented these methods based on their public code and our DG settings.
	Tab.~\ref{tab:CompSS} shows the performance comparison with the methods.
	\textbf{SleepDG achieves the state-of-the-art performance in all the datasets} (75.03\% in average ACC and 69.64\% in average MF1), proving that SleepDG can better solve the generalization problem of unseen domain compared with traditional sleep staging methods even if subject population and signal channels of source domains are different from those of the unseen domain.
	The DeepSleepNet performs the worst, about 7.6\% lower in average ACC and 7.7\% lower in average MF1 than SleepDG, indicating that classical model is difficult to deal with DG scenarios.
	TinySleepNet performs a little better than DeepSleepNet (68.49\% v.s. 67.48\% in average ACC and 62.60\% v.s. 61.93\% in average MF1).
	It indicates that a smaller number of model parameters may alleviate overfitting issue in DG scenarios.
	U-Time, AttnSleep, ResnetMHA and EnhancingCE cannot obtain a significant improvement in performance compared with the classical method DeepSleepNet, indicating that it is difficult to achieve a significant performance improvement in DG scenarios by only modifying the network structures.
	SalientSleepNet achieves a significant performance improvement on specific dataset~\citep{Jia2021SalientSleepNetMS}, but it only performs about 1.5\% higher in average ACC and 1.6\% higher in average MF1 compared with DeepSleepNet, indicating that a complex neural network design cannot solve the problem of domain shift well.
	Compared with these traditional non-DG methods, SleepDG can learn domain-invariant features through domain alignment, and further improve generalization ability to unseen domains.

	\subsubsection{Compared with DG Methods for Sleep Staging}
	\label{sec:dg}

	We compared SleepDG with BASE and other classical DG methods in sleep staging:
	\textbf{BASE} can be regarded as SleepDG which minimizes $\mathcal L_{\mathrm{classify}}$ but without any reconstruction and DG loss.
	\textbf{MMD} (Maximum Mean Discrepancy) \citep{tzeng2014deep, li2018domain} is a DG method matching the MMD of feature distributions.
	\textbf{CORAL} \citep{sun2016deep} matches the covariance of feature distributions.
	\textbf{IRM} (Invariant Risk Minimization) \citep{Arjovsky2019InvariantRM} enforces the optimal classifier to be the same across all domains.
	\textbf{DANN} (Domain-Adversarial Neural Networks) \citep{ganin2015unsupervised} employs an adversarial network to match feature distributions.
	
	We implemented these DG methods based on BASE and DeepDG toolkit~\citep{wang2022generalizing}, and the performance comparison is shown in Tab.~\ref{tab:CompDG}.
	SleepDG achieves the best performance compared with other DG methods.
	BASE performs the worst (68.76\% in average ACC and 63.34\% in average MF1), indicating that a model without DG loss cannot minimize the domain shift among different domains.
	MMD performs better compared with BASE, about 3.0\% higher in average ACC and 2.7\% higher in average MF1, suggesting minimizing maximum mean discrepancy can extract domain-invariant features to a degree.
	CORAL performs the best in existing DG methods (72.27\% in average ACC and 67.52\% in average MF1), which indicates that feature alignment based on second-order statistics (covariance) can achieve good cross-domain generalization.
	IRM performs better than MMD but worse than CORAL (72.09\% in average ACC and 66.95\% in average), indicating that invariant risk minimization has no obvious advantage over feature alignment methods.
	The domain-adversarial model of DANN achieves the least improvement compared with BASE (71.12\% v.s. 68.76\% in average ACC).
	It matches the finding in \citet{wang2022generalizing}: although domain adversarial training often achieves better performance in DA, it is difficult to make a significant improvement in DG.
	Compared with these classical DG methods, SleepDG can learn domain-invariant features better by combining Epoch-level Feature Alignment and Sequence-level Feature Alignment.
	

	\subsubsection{Ablation Study}
	
	To investigate the effectiveness of signal reconstruction and feature alignment in SleepDG, we conducted an ablation study.
	Here, we also take \textbf{BASE} as our baseline model.
	The ablated methods are
	\textbf{AE} (Base + Autoencoder),
	\textbf{EA} (Base + Epoch-level Feature Alignment),
	\textbf{SA} (Base + Sequence-level Feature Alignment),
	\textbf{AE+EA} (Base + Autoencoder + Epoch-level Feature Alignment)
	and \textbf{AE+SA} (Base + Autoencoder + Sequence-level Feature Alignment).
	The results are shown in Tab.~\ref{tab:AblationStudy}. AE performs a little better than BASE about 0.97\% higher in average ACC and 1.09\% higher in average MF1, indicating that AE-based representation learning can learn generalizable features to a certain extent.
	EA contributes greatly to the sleep staging, obtaining a great performance improvement compared with BASE, 3.57\% and 4.07\% higher in average ACC and average MF1, respectively.
	AE+EA also performs well, obtaining 4.42\% and 4.71\% higher in average ACC and average MF1 than Base.
	It indicates the Epoch-level Feature Alignment can extract domain-invariant epoch features well.
	Meanwhile, compared with Base, SA performs about 2.37\% higher in average ACC and 3.16\% higher in average MF1, and AE+SA performs about 3.10\% and 3.85\% higher, indicating that Sequence-level Feature Alignment can align context features.
	SleepDG can further improve the ability to learning domain-invariant features in both epoch and sequence level, improving the performance of 6.27\% in average ACC and 6.30\% in average MF1 compared with BASE.
	Notably, SleepDG achieves the best performance across all the unseen domain settings, indicating that SleepDG has a relatively stable generalization performance.

	\subsubsection{Feature Visualization}

\begin{figure}[tb]
	\centering
	\small
	\includegraphics[width=0.98\columnwidth]{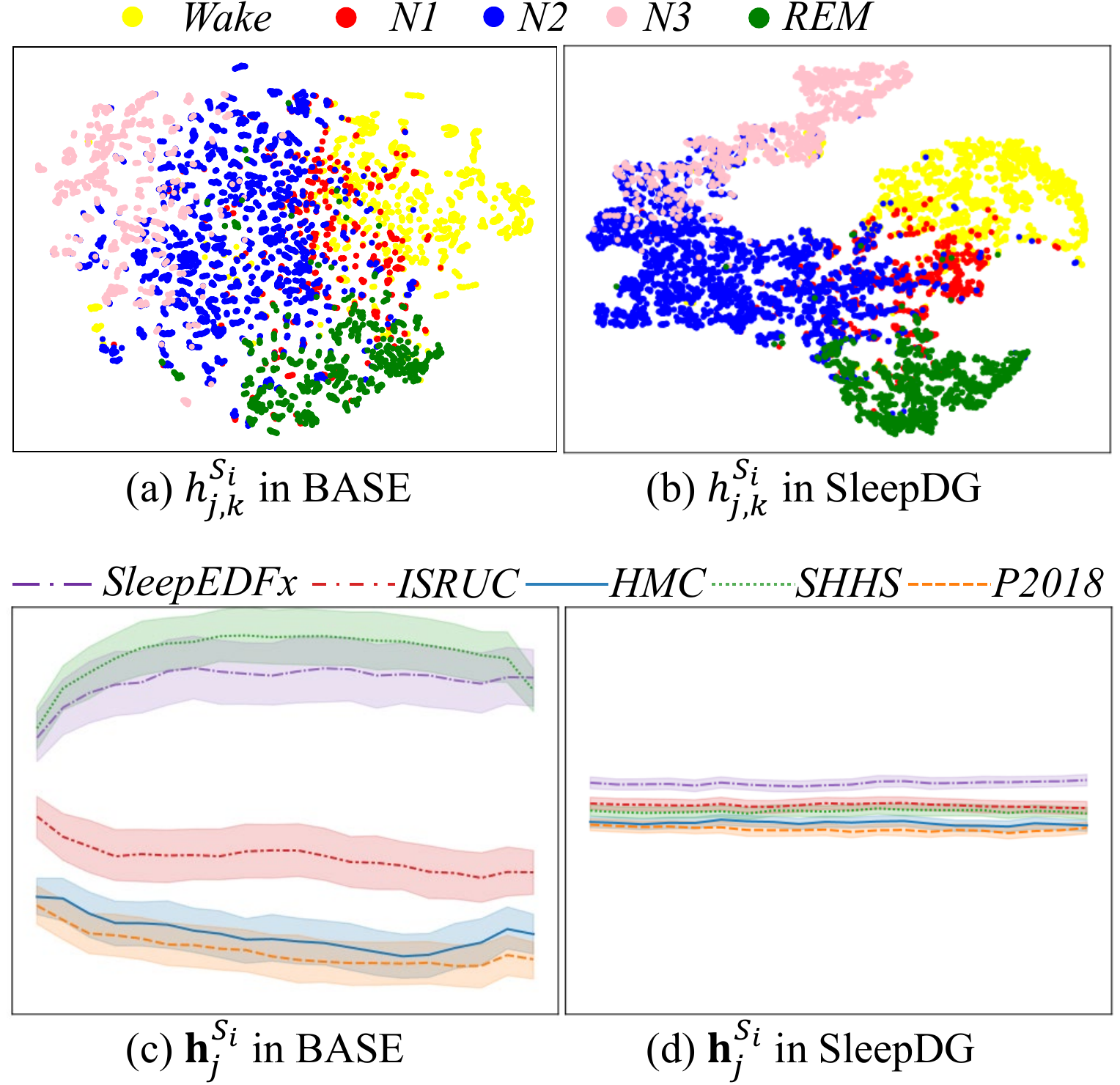}
	\caption{Feature visualization of BASE and SleepDG.}
	\label{fig:Case}
\end{figure}

	We conducted a feature visualization to further show the effectiveness of SleepDG as Fig.~\ref{fig:Case} shows.
	Fig.~\ref{fig:Case}(a)(b) show epoch features visualization of different domains from BASE and SleepDG. The different colors represent different sleep stages.
	Here, we visualize the features by t-SNE \citep{van2008visualizing}.
	Compared with the feature representations in Fig.~\ref{fig:Case}(a), in Fig.~\ref{fig:Case}(b) the feature representations in the same sleep stage obviously form a cluster and the ones in different sleep stages nicely separate with each other after we adopt our alignment methods, even though these features are totally from different domains. It intuitively indicates that our SleepDG can learn domain-invariant feature representations which are independent on the specific domains.
	Meanwhile, it further proves that the feature representations obtained by SleepDG are more distinguishable for automatic sleep staging.
	Fig.~\ref{fig:Case}(c)(d) show sequence feature visualization of BASE and SleepDG.
	We use PCA to reduce the dimension of every $h^{S_i}_{j, k}$ in sequence features $\mathrm{\mathbf{h}}^{S_i}_j$ to one dimension and draw a linear regression line plot for each domain.
	Obviously, the sequence features from SleepDG form a closer cluster compared with those from BASE, which intuitively indicates that our Sequence-level Feature Alignment is effective to align the inter-epoch relationship.
	
	\section{Conclusion}
	We propose a novel task of generalizable sleep staging to solve the severe generalization problem of automatic sleep staging in clinic. In order to improve the generalization ability of sleep staging methods, we propose a novel DG-based framework. Considering both of the local salient features within each sleep epoch and sequential features among different epochs, we develop a Multi-level Feature Alignment method which focuses on both epoch-level and sequence-level feature alignment. Specifically, we design an Epoch-level Feature Alignment method aligning both mean and covariance of single-epoch feature distribution among different domains to learn epoch-level domain-invariant features. We also design a Sequence-level Feature Alignment method to extract sequence-level domain-invariant features, by minimizing the discrepancy of Pearson correlation coefficients of sleep sequence among different domains. 
	Our method is validated on five public sleep datasets with DG settings, achieving the state-of-the-art performance.
	
	\section{Acknowledgments}
	This work was supported by STI 2030 Major Projects (2021ZD0200400), Natural Science Foundation of China (No. 61925603) and the Key Program of the Natural Science Foundation of Zhejiang Province, China (No. LZ24F020004).
	The corresponding authors are Dr. Sha  Zhao and Dr. Gang Pan.
	\bibliography{GSS-references}
	
\end{document}